\documentclass[aps,amsfonts,amsmath,prd,preprint,nofootinbib]{revtex4}
\usepackage{xcolor}  

\newcommand{\beq}{\begin{equation}}
\newcommand{\eeq}{\end{equation}}

\def\beq{\begin{equation}}
\def\eeq{\end{equation}}
\def\beqa{\begin{eqnarray}}
\def\eeqa{\end{eqnarray}}

\usepackage{epsfig}

\begin{document}

\title{Quantum instability of an oscillating universe}

\author{Thibault \surname{Damour}$^1$ and Alexander \surname{Vilenkin}$^2$}

\affiliation{
$^1$Institut des Hautes Etudes Scientifiques, 91440 Bures-sur-Yvette, France\\
$^2$ Institute of Cosmology, Department of Physics and Astronomy,\\ 
Tufts University, Medford, MA 02155, USA}

\begin{abstract}

An oscillating, compact Friedmann universe with a massive conformally coupled scalar field is studied in the framework of quantum cosmology.  The scalar field is treated as a perturbation and we look for solutions of the Wheeler-DeWitt equation describing stable stationary states of the model.  We assume that the previous sources of quantum instability that have been discussed in the
literature (particle production, and tunnelling to zero size) are absent. We then show, under rather general assumptions, that 
a further source of quantum instability prevents the existence of stationary states with localized wave function in the direction of
the scalar-field modes.  

\end{abstract}

\maketitle

\section{Introduction}

In recent years there has been much interest in the possible existence of absolutely stable models describing a universe that exists forever in a static or oscillating state.  Apart from the intrinsic interest of such models, they can play the role of an `eternal seed', providing a starting point for the emergent universe scenario (see, e.g, \cite{Mulryne:2005ef} and references therein).

Construction of classical oscillating cosmologies is relatively straightforward.  See Ref.~\cite{Dabrowski1}
for a  general class of oscillating models. A particularly simple oscillating model of a closed Friedmann universe was discussed in 
Ref.~\cite{Graham:2011nb}: it includes a {\it negative} cosmological constant $\Lambda $ and a matter component having the effective equation of state $P=w\rho$ with $w=-2/3$ .  A  negative $\Lambda <0$ causes an expanding universe to turn around 
and recollapse, while the $w=-2/3$ matter component provides an energy density $\rho_w=\rho_0 a^{-3(1+w)}= \rho_0 a^{-1} $  which causes a collapsing universe to bounce.  The evolution equation for this model has solutions where the scale factor changes sinusoidally with time, so the model has been dubbed ``a simple harmonic universe" (SHU).  
We will give below another example of matter content leading to a simple sinusoidal variation of the square of the scale factor.

In such models the universe would classically oscillate forever between two scale factor values $a_1$ and $a_2$, but one needs to watch for potential quantum instabilities.  A first source of instability is that the oscillation of the universe can induce particle 
production, which would be manifested by exponentially growing modes of quantum fields.  The analysis in 
Refs.~\cite{MV2,Graham2} has shown that unstable modes are indeed present for a significant part of the parameter space in a number of oscillating models, including SHU.  However, there remains a significant, complementary range of the parameter 
space where the models do not exhibit any unstable modes.  We shall assume here that we are in a parameter range 
where our oscillating model is stable under particle production. A second possible source of quantum instability is non perturbative:
 it was argued in Refs.~\cite{MV1} (following the earlier work of \cite{Dabrowski}) that oscillating models are generically unstable with respect to collapse to zero volume via quantum tunneling.  However, the presence or absence
 of this instability is linked to the choice of a boundary condition, at zero volume, for the wave function $\Psi$ of the universe, an
 issue which concerns the deep UV regime of quantum gravity. To fix ideas, we shall assume below
 (following an old suggestion of DeWitt \cite{DeWitt}) that $\Psi$ vanishes on singular universes, thereby avoiding the
 instability under tunneling to zero size\footnote{We note in passing that, in the quantum supergravity model of Ref. \cite{Damour}, 
 there exist exact quantum states describing seed universes where  $\Psi$ vanishes (without fine-tuning) both for small
 and large volumes.}. However, we shall see that this assumption is not important for reaching our main result.

In the present paper we shall study an oscillating universe in the framework of quantum cosmology, where both the scale factor $a$ and the quantum field modes $\phi_n$ are described by a wave function $\Psi(a,\phi_n)$.  
This wave function is defined not only in the classical range $a_1<a<a_2$, but also extends beyond it into the classically forbidden regions.  A stable oscillating universe would correspond to a stationary state described by a wave function which is localized 
around the classical solution (corresponding to a periodic function $a(t)$ oscillating between $a_1$ and $a_2$, and to
the classical ground state $\phi_n=0$), and which exponentially decreases away from this solution,  both in the $a$ and $\phi_n$ directions.  Our main result is that such stationary states generally do not exist. 

\section{Wheeler-DeWitt equation}

We consider a {\it compact} ($k=+1$) Friedmann universe filled with homogeneous matter of energy density $\rho(a)$, and with a massive 
conformally coupled scalar field $\phi({\bf x},t)$, whose background value vanishes, and which is treated as a perturbation
at the quantum level.\footnote{A minimally coupled field can be analyzed in a similar way with essentially the same results.}  The precise form of the function $\rho(a)$ will not be important for our analysis.  We expand the field $\phi$ as
\beq
\phi({\bf x},t)=\frac{1}{a(t)}\sum_n \phi_n(t) Q_n({\bf x}),
\eeq
where $Q_{nlr}({\bf x})$ are suitably normalized harmonics on the 3-sphere, $n=1,2, ...$. We have suppressed the indices $l,r$ for brevity.  After appropriate rescalings (see, {\it e.g.}, \cite{Vilenkin:1987kf}), the Hamiltonian of this model can be written as
\beq
{\cal H}=-p_a^2 -U(a)+\sum_n \left(p_n^2+\omega_n^2(a) \phi_n^2\right),
\eeq
where $p_a$ and $p_n$ are momenta conjugate to $a$ and $\phi_n$ respectively,
\beq
U(a)=a^2 \left(k- a^2 \rho(a)\right),
\eeq
\beq
\omega_n^2(a)=n^2+m^2 a^2,
\eeq
$k$ is the curvature parameter (here taken to be +1) and $m$ is the mass of the field $\phi$. 

In quantum theory the momenta are replaced by differential operators, $p_a\to -i\hbar\partial/\partial a$, $p_n\to -i\hbar\partial/\partial \phi_n$, and the classical Hamiltonian constraint ${\cal H}=0$ is replaced by the Wheeler-DeWitt (WDW) equation for the wave function of the universe,
\beq
{\cal H}\Psi(a,\{\phi_n\})=0,
\eeq
or explicitly
\beq
\left[\hbar^2\frac{\partial^2}{\partial a^2} -U(a) +\sum_n\left(-\hbar^2\frac{\partial^2}{\partial \phi_n^2} +\omega_n^2(a) \phi_n^2 \right)\right] \Psi =0.
\label{WDW}
\eeq
Here we write the Planck constant $\hbar\equiv 1$ only as a formal bookkeeping parameter, to clarify the use of the WKB approximation.

The probabilistic interpretation of the wave function $\Psi$ is a matter of debate.  Some authors \cite{interpretation,Halliwell} 
have advocated that the probability should be expressed in terms of the conserved current,
\beq \label{J}
J=i(\Psi^*\nabla\Psi-\Psi\nabla\Psi^*),
\eeq  
where the gradient $\nabla$ is with respect to the superspace variables $\{a,\phi_n\}$.  This approach has a number of attractive properties, but it fails in the case of a stationary state, where $J=0$.  An alternative approach \cite{HP1} is to use the `naive Schr\"odinger measure',
\beq
{\cal P}\propto |\Psi|^2.
\label{naive}
\eeq
It has been argued in \cite{HP1} that the two approaches agree under certain assumptions.  We will not attempt to resolve this issue here.
In the present work, it will be enough to argue from the quadratic-in-$\Psi$ structure of both expressions, Eqs. \eqref{J} and \eqref{naive},
that $\Psi$ should be constrained to tend to zero far away from the support of the classical solution.  

In particular, as our model is intended to describe the quantum version of a
classical solution where  a periodic scale factor $a^{\rm classical}(t)$ oscillates between $a_1$ and $a_2$, and where
 $\phi^{\rm classical}({\bf x},t)=0$,  we should require that the wave function $\Psi(a, \phi_n)$ satisfy
 \beq
\lim_{a\to\infty} \Psi(a, \phi_n)=0 \,,
\label{bcinfty}
\eeq
and
\beq
\lim_{\phi_n\to \pm \infty}\Psi(a, \phi_n)=0\,.
\eeq

As already mentioned, we shall also tentatively require that $\Psi$ vanishes near zero-volume, singular universes:
\beq
\lim_{a\to0} \Psi(a, \phi_n)=0 .
\label{bc0}
\eeq
It is possible that a UV-complete theory of gravity might resolve the singularity at $a=0$, replacing it with a non-singular Planck-size nugget.  The universe may then tunnel back and forth between the semiclassical oscillating regime and the nugget, resulting in a stationary quantum state.  Eq.~(\ref{bc0}) would then be an approximate boundary condition, requiring that the probability of finding the universe at $a\lesssim 1$ is very small.

Using a standard technique \cite{Banks,Lapchinsky,HH}, an approximate semiclassical solution to the WDW equation can be 
constructed as a linear combination of $\phi_n$-Gaussian terms of the form
\beq
\Psi (a,\{\phi_n\})=\exp\left[-\frac{S(a)}{\hbar}-\frac{1}{2\hbar}\sum_n S_n(a) \phi_n^2\right],
\label{Gaussian}
\eeq
where $S(a)$ and $S_n(a)$ are {\it complex} functions of $a$, to be determined.  For a stationary state the functions $S_n(a)$ must satisfy the stability condition \cite{Vachaspati},
\beq
{\rm Re}\left[ S_n(a) \right] > 0,
\label{regularity}
\eeq
which ensures that the modulus of the wave function decreases with increasing amplitude of the field fluctuations $\phi_n$. 

Substituting the ansatz of Eq.~(\ref{Gaussian}) into the WDW equation (\ref{WDW}) we obtain the conditions
\beq \label{WDWa}
{S'}^2-U(a)-\hbar S'' +\hbar\sum_n S_n=0,
\eeq
\beq \label{WDWphi}
\left[S'{S_n}' -S_n^2+\omega_n^2(a) -\frac{\hbar}{2}{S_n}'' \right] \phi_n^2 +\frac{1}{4} {{S_n}'}^2 \phi_n^4 = 0.
\eeq
Our approximation consists of disregarding the terms of order ${\cal O}(\hbar)$ and ${\cal O}(\phi_n^4)$; a detailed discussion of its applicability can be found in Ref.~\cite{Hong}.  We then have the following WKB equations for the functions $S(a)$ and $S_n(a)$,
\beq
{S'}^2-U(a)=0,
\label{Seq}
\eeq
\beq
S' {S_n}' -S_n^2+\omega_n^2=0.
\label{Sneq}
\eeq
From now on we set $\hbar=1$.

\section{Solving the WDW equation for the quantum dynamics of the scale factor: $\Psi_0(a)$}

We first consider the part $\Psi_0(a) \equiv \exp\left[- \frac{1}{\hbar}S(a)  \right]$ of the wave function that depends
only on the scale factor. We will add later the $\phi_n$-dependent factor 
$\exp\left[-\frac{1}{2\hbar}\sum_n S_n(a) \phi_n^2\right]$.

 In our notation, an oscillating universe corresponds to a WDW potential $U(a)$ that is {\it negative}, $U(a)<0$, in 
the oscillation range $a_1 <a<a_2$, and positive, $U(a)>0$, in the classically forbidden regions on both sides of this range.  
 The classical version of this model admits solutions where $a(t)$ oscillates periodically between the turning points $a_1$ and $a_2$.

The solutions of the zeroth order equation (\ref{Seq}) in the classically allowed range are
\beq
S^{\pm}(a) = \pm i\int_{a}^{a_2} \sqrt{-U(a')} da'
\label{allowed}
\eeq
with upper and lower signs corresponding to contracting and expanding universe, respectively.  For a stationary state the $a$-part
of the wave function should be a superposition of such terms,\footnote{The standard WKB pre-exponential factors $\propto [-U(a)]^{-1/4}$ can be determined by solving the WDW equation to second order in $\hbar$.  These factors do not affect our conclusions and we disregard them here.} 
\beq
\Psi_0(a) = A_+ e^{-S^+(a)}+A_- e^{-S^-(a)}=  A_+ e^{- i \int_{a}^{a_2} \sqrt{-U(a')} da'}+A_- e^{+i \int_{a}^{a_2} \sqrt{-U(a')} da'} ~~~~ (a_1 < a < a_2)\,,
\label{psi0}
\eeq   
with $A_\pm = {\rm const}$.
In the range $a>a_2$ we have exponentially growing and decreasing solutions $\exp{ \pm {\bar S}_{2}^+(a)}$ with
\beq
{\bar S}_{2}^+(a) \equiv + \int_{a_2}^{a} \sqrt{U(a')} da' > 0, ~~~~ (a>a_2)\,,
\label{forbidden}
\eeq
where the bar over $S$ indicates that the solution is in a classically forbidden region, and the index $2$ that we are in the range beyond $a_2$.
They can be matched to those in the allowed region using the WKB connection formulas \cite{LandauLifshitzQM}.  
The boundary condition (\ref{bcinfty}) selects 
the decreasing solution   
\beq
\Psi_0(a)=B_2 e^{-{\bar S}_{2}^+(a)} ~~~~ (a>a_2)\,,
\eeq
and the WKB connection formulas give the relations
\beq
B_2=e^{-i\pi/4} A_+ = e^{i\pi/4} A_-.
\eeq

In the second classically forbidden region $0\leq a < a_1$ we also have exponentially growing and decreasing solutions,
$\exp{ \pm {\bar S}_{1}^+(a)}$, with
\beq
{\bar S}_{1}^+ \equiv + \int_{a}^{a_1} \sqrt{U(a')} da'  > 0, ~~~~ (0< a < a_1),
\label{forbidden2}
\eeq
and the boundary condition (\ref{bc0}) can be approximately implemented by selecting
\beq
\Psi_0(a)=B_1 e^{-{\bar S}_{1}^+(a)} ~~~~ (0<a<a_1)\,.
\eeq
The matching conditions at $a=a_1$ then require
\beq
B_1 =A_+ e^{i\pi/4} e^{-iQ}= A_- e^{-i\pi/4} e^{iQ},
\eeq
where 
\beq
Q=\int_{a_1}^{a_2} \sqrt{-U(a)} da.
\eeq
It follows that $Q$ must be equal to
\beq
Q=\left(n+\frac{1}{2}\right)\pi,
\label{Q}
\eeq
where $n$ is an integer, and that
\beq
B_1=e^{i\pi/4} A_- = B_2 .
\eeq

Eq.~(\ref{Q}) is just the Bohr-Sommerfeld quantization condition. It must be satisfied because
of our double boundary condition \eqref{bc0}, \eqref{bcinfty}. One way to enforce it (if it is not automatically
satisfied as in the quantum supergravity model of Ref. \cite{Damour}) is to fine-tune the matter content of the universe.  For example, one can introduce an adjustable amount of radiation with energy density $\rho_r=\epsilon_r a^{-4}$, where $\epsilon_r={\rm const}$.  The WDW potential is then replaced by $U(a)-\epsilon_r$ and the WDW equation for $\Psi_0(a)$ becomes
\beq
\left[-\frac{d^2}{da^2}+U(a)\right]\Psi_0=\epsilon_r\Psi_0.
\eeq
This has the same form as the standard energy eigenvalue equation, and the condition (\ref{Q}) determines the spectrum of allowed values of $\epsilon_r$.\footnote{In our semiclassical treatment the quantity $\epsilon_r$ characterizing the amount of radiation is assumed to be a continuous parameter.  It is not clear that such continuously tunable parameters will exist in the full quantum theory.  One might expect that, on the contrary, $\epsilon_r$ could be quantized.  Then there is no guarantee that its spectrum will allow us to enforce our boundary conditions.}

\section{Completing the solution with $\phi_n$-dependent factors}

The full wave function, including scalar field perturbations, can be written as
\beq
\Psi(a,\{\phi_n\}) = A_+ \exp\left[-S^+(a) -\frac{1}{2}S_n^+(a) \phi_n^2 \right] 
+ A_- \exp\left[-S^-(a) -\frac{1}{2}S_n^-(a) \phi_n^2 \right]                 ~~~~ (a_1 < a < a_2)
\label{psiallowed}
\eeq   
in the classically allowed region and
\beq
\Psi(a,\{\phi_n\}) = B_{\{1,2\}} \exp\left[-{\bar S}_{\{1,2\}}^+(a) -\frac{1}{2}{\bar S}_{n\{1,2\}} (a) \phi_n^2 \right] ~~~~ (a<a_1,~ a>a_2)
\eeq
in the two classically forbidden regions.  The functions $S_n^\pm(a)$ and ${\bar S}_{n\{1,2\}} (a)$ satisfy Eq.~(\ref{Sneq}) with $S(a)=S^\pm(a)$ and $S(a)= {\bar S}_{\{1,2\}}^+(a)$, respectively.  Continuity requires that these functions have to match at the turning points $a_1$ and $a_2$ \cite{Vachaspati}:
\beq
S_n^+(a_1)=S_n^-(a_1)={\bar S}_{n1}(a_1),
\label{matching1}
\eeq
\beq
S_n^+(a_2)=S_n^-(a_2)={\bar S}_{n2}(a_2).
\label{matching2} 
\eeq
The matching conditions \eqref{matching1}, \eqref{matching2}, will play a central role in our reasonings.

\subsection{Uniqueness of the solution for $S_n(a)$ in the outer classically forbidden region $a>a_2$}

We start our analysis by considering  the outer classically forbidden region, $a>a_2$. In this region
 $S(a) ={\bar S}_{2}^+(a) \equiv + \int_{a_2}^{a} \sqrt{U(a')} da' > 0$ is real
(and growing with $a$). It is then convenient to define an auxiliary Euclidean time variable $\tau$ by
\beq 
\frac{da}{d\tau}=S'(a)=+ \sqrt{U(a)} ~~~~ (a>a_2).
\label{atau}
\eeq
This Euclidean time grows with $a (>a_2)$, and can be set to be zero when $a=a_2$.

Then Eq.~(\ref{Sneq}) becomes
\beq  \label{riccatiforbidden}
{\dot S}_n-S_n^2+\omega_n^2=0,
\eeq
where a dot stands for a derivative with respect to $\tau$.  

Let us first note that Eq. \eqref{riccatiforbidden} is a first-order ordinary differential equation (ODE). Its general solution
will therefore contain only one arbitrary constant. We must impose on the solution $S_n(\tau)$ to 
satisfy the stability inequality \eqref{regularity}. We are going to show that it is possible to satisfy the inequality (\ref{regularity})  
in the entire region $a>a_2$, but that this condition {\it uniquely} determines a solution of \eqref{riccatiforbidden}. 
This surprisingly strong effect of
imposing the apparently benign sign inequality \eqref{regularity} is due to the exponential behavior of the generic solution
of the Euclidean-time equation \eqref{untaueq}.

Eq. \eqref{riccatiforbidden} is a nonlinear ODE of the Riccati type. As is well-known such a Riccati equation can be transformed
into a {\it linear} second-order ODE by representing $S_n$ in the form of a logarithmic
derivative:
\beq   \label{Snuntau}
S_n=-\frac{{\dot u}_n}{u_n}\,.
\eeq
Then $u_n(\tau)$ must satisfy the linear equation
\beq
{\ddot u}_n =+\omega_n^2(\tau) \,u_n .
\label{untaueq}
\eeq

The general solution of the second-order ODE \eqref{untaueq} contains (for each mode number $n$) {\it two} arbitrary
constants, say $C_{n +}$ and   $C_{n -} $. However, the logarithmic derivative $ \dot u_n/u_n$ parametrizing $S_n(\tau)$ 
only depends on the ratio  $C_{n +} /C_{n -}$ so that solving \eqref{untaueq} leads to the same number (namely one) of arbitrary
constants as solving the original equation, Eq. \eqref{riccatiforbidden}, for $S_n(\tau)$.
We look for a solution of \eqref{untaueq}  such that $S_n=-{\dot u}_n/u_n$ satisfies the inequality (\ref{regularity}). 

Let us first discuss the asymptotic behavior of the mode functions at large $a$ by approximating the general solution
of Eq. \eqref{untaueq} by means of the WKB approximation,
\beq
u_n(\tau)\approx  C_{n +} \exp\left[\int^\tau_0 \omega_n(\tau')d\tau'\right] + C_{n -} \exp\left[-\int^\tau_0 \omega_n(\tau')d\tau'\right]\,,
\label{untau}
\eeq
where $C_{n \pm} $ are two {\it a priori} complex constants. 
We shall later indicate that the results derived from the latter WKB approximation can
be confirmed by using corresponding exact solutions of Eq. \eqref{untaueq}.

The integral in the exponents read
\beq
\int^\tau_0 \omega_n(\tau')d\tau' =\int^a_{a_2} da' \frac{ \sqrt{n^2+ m^2 a'^2}}{\sqrt{U(a')}}.
\eeq
Its asymptotic behavior for large values of $a$ can be expressed as
\beq
\int^\tau \omega_n(\tau')d\tau' \sim m \int^a \frac{a' da'}{\sqrt{U(a')}}.
\eeq  
For all of the known (and most of the hypothesized) forms of matter, the energy density $\rho(a)$ decreases with $a$ or at most remains constant.  Then $U(a)$ does not increase faster than $a^4$ and the above integral diverges at $a\to\infty$.   
Hence, if $C_{n +}\neq 0$, the first term in (\ref{untau}) dominates at large $a$ and the representation \eqref{Snuntau} yields
an asymptotically real $S_n(a)$ satisfying
\beq
S_n(a\to\infty) \approx -\omega_n(a) \sim  - m a  \to - \infty ~~~~ ({\rm if} ~~ C_{n +}\neq 0).
\eeq
Here, the crucial minus sign in front of $\omega_n(a)$ (which implies that $S_n(a)$ is generically negative for large $a$)
comes from the minus sign in front of ${\dot u}_n/u_n$ in
Eq. \eqref{Snuntau}, and the latter minus sign was uniquely determined by the various signs in the Riccati equation  
\eqref{riccatiforbidden}.

To avoid violation of the regularity condition \eqref{regularity}, we should therefore set $C_{n +}=0$, so that
\beq
S_n(a) \approx +{\omega_n(a)} > 0 ~~~~ ({\rm if} ~~ C_{n +} = 0).
\eeq

Therefore, within the WKB approximation \eqref{untau} for $u_n(\tau)$, the choice  $C_{n +} = 0$ (for all $n$'s) ensures that
the regularity condition \eqref{regularity} is satisfied in the whole region $a>a_2$.  Let us actually show that
this result remains true without making use of the WKB approximation to solve the linear ODE
 \eqref{untaueq}. Indeed, let us first remark that Eq. \eqref{untaueq} is a real equation,
which therefore admits two independent {\it real} solutions. Actually, Eq. \eqref{untaueq} has the general form of
Schr\"odinger's (fixed energy) one-dimensional equation in a classically forbidden interval. It is shown in Messiah's Quantum
Mechanics treatise \cite{Messiah} (using Floquet-theory-type reasonings) that among the {\it real} solutions normalized to
satisfy $u_n(\tau=0)=1$ there is a {\it unique} one, say $u_n^-(\tau)$, such that 
\beq
u_n^-(\tau)>0, ~~~{\rm and} ~~~ {\dot u}_n^-(\tau)<0\,,
\eeq
on the entire interval of variation of $\tau$, with $u_n^-(\tau) \to 0$ as $\tau$ tends to the upper limit of its
interval of variation\footnote{$\tau$ varies between 0 and $\tau_m=\int_{a_2}^{\infty} da/\sqrt{U(a)}$. Depending
on the oscillating-universe model, $\tau_m$ can be infinite or finite. However, what is crucial for our results is that
the total ``number of e-folds" $\int  \omega_n d\tau = \int^{\infty}_{a_2} da  \sqrt{n^2+ m^2 a^2}/\sqrt{U(a)}$
be infinite, which is the case in all models, as we already pointed out.}. As second independent (real) solution one can take
the solution $u_n^+(\tau)$ with initial data $u_n^+(\tau=0)=1$ and ${\dot u}_n^+(\tau=0)=0$. The latter solution 
is similar to the WKB solution $\cosh\left[\int^\tau_0 \omega_n(\tau')d\tau'\right] $: it
stays positive,  grows monotonically (and exponentially), as $\tau \to \tau_m$, and is such that
\beq
\frac{{\dot u}_n^+(\tau)}{u_n^+(\tau)} \sim +\omega_n(\tau) \to + \infty ~~~ {\rm as} ~~~ \tau\to \tau_m\,.
\eeq
Redoing our reasoning above with the exact general solution
\beq
u_n(\tau) =  C_{n +} \, u_n^+(\tau) + C_{n -} \, u_n^-(\tau),
\eeq
leads to the same conclusions: if $C_{n +}\neq 0$, we have $S_n(a) = - \dot u_n/u_n \sim -  \omega_n(a)  \to - \infty $
as $a \to \infty$. The only stable solution is obtained for $C_{n +}= 0$, and it satisfies  
\beq \label{Sn2unique}
\left[ {\bar S}_n^{\rm outer}(a)\right]_{C_{n +}= 0} = - \frac{\dot u_n^-(\tau)}{u_n^- (\tau)} >0. 
\eeq

Summarizing: we have shown that there is a unique solution of Eq. \eqref{riccatiforbidden} in the outer
classically forbidden range $a >a_2$ that is physically acceptable (namely  satisfying the  condition \eqref{regularity}). It is given by Eq. \eqref{Sn2unique}. Note that this unique solution is real, and therefore satisfies not only ${\rm Re}[S_n(a)] >0$,
but actually the stronger condition $S_n(a) >0$.

In view of our matching condition \eqref{matching2}, the above constructed unique stable solution
yields, for each mode number $n$, a specific boundary condition for $S^{\pm}_n(a)$ at their outer boundary, say
\beq \label{matching3}
S_n^+(a_2)=S_n^-(a_2)= \mu_n >0\,,
\eeq
where we defined
\beq
\mu_n \equiv  - \frac{\dot u_n^-(0)}{u_n^- (0)} =\left[ {\bar S}_n^{\rm outer}(a)\right]_{C_{n +}= 0}\,.
\eeq
Here, we have introduced the notation $\mu_n$ for the uniquely defined sequence of (minus) the logarithmic derivatives
of $u_n^- (\tau)$, evaluated at $\tau=0$. The real, positive numbers $\mu_n$ are characteristic numbers associated with
the functions $\omega^2_n(\tau)$ (considered in the outer forbidden region). They will play a crucial role in the rest of
our analysis.

\subsection{Non-existence of a solution for $S^{\pm}_n(a)$ in the classically allowed region matching the 
outer solution ${\bar S}^{\rm outer}_{n}(a)$}

In the classically allowed range, where ${S}'(a)$ is pure imaginary, it will be convenient to introduce an 
auxiliary (Minkowskian) time variable $t$ which is related to $a$ by the classical equation of motion
\beq 
\frac{da}{dt}=-i{S}'(a) =\pm \sqrt{-U(a)}.
\label{xt}
\eeq
Then Eq.~(\ref{Sneq}) becomes
\beq \label{riccatiallowed}
-i{\dot S}_n +S_n^2-\omega_n^2=0,
\eeq
where a dot now stands for a derivative with respect to $t$.

Eq. \eqref{riccatiallowed} is again a nonlinear ODE of the Riccati type (which is now complex). We  transform it
into a {\it linear} second-order ODE by representing $S_n$ as the following (complex) logarithmic
derivative:
\beq \label{Snvsun}
S_n=-i \frac{\dot u_n}{u_n}.
\eeq
Then $u_n(t)$ must satisfy the linear equation
\beq
{\ddot u}_n + \omega_n^2 [a(t)] u_n =0,
\label{Hill}
\eeq
with $a(t)$ satisfying Eq.~(\ref{xt}). Note that this is the Minkowskian-time version ($ t \to i \tau$) of the corresponding 
forbidden-range  equation \eqref{untaueq}. As in the previous subsection, though
the general solution of the second-order ODE \eqref{Hill} contains {\it two} arbitrary
constants, say ${ C}_n$ and   ${\tilde C}_n $, the logarithmic derivative $ \dot u_n/u_n$ parametrizing $S_n(t)$ 
only depends on the ratio  ${\tilde C}_n /C_n$ so that solving \eqref{Hill} does not introduce more arbitrary
constants than solving the original Riccati equation, Eq. \eqref{riccatiallowed}, for $S_n(t)$.

The upper choice of sign in Eq.~(\ref{xt}) corresponds to an expanding universe.  The corresponding classical solution $a(t)$ varies from $a_1$ to $a_2$ with ${\dot a}=0$ at both ends, and its quantum-mechanical counterpart is the first term in Eq.~(\ref{psiallowed}).  For the lower choice of sign, the universe contracts back from $a_2$ to $a_1$.  Combining the two solutions we obtain a periodic function $a(t)$,
of period $T$: $a(t+T)= a(t)$.  Eq.~(\ref{Hill}) (which happens to coincide with the classical  equation for scalar field modes in the oscillating universe background)
  is a Hill equation, and  the stability of its solutions can be analyzed using Floquet theory \cite{Floquet}.

We recall that Floquet theory consists in expressing the general solution of Eq.~(\ref{Hill}) as a linear combination
of the two eigenmode solutions of Eq.~(\ref{Hill}) that reproduce themselves, modulo a multiplicative factor $\lambda_n^{\pm}$,
after a period $T$. The eigenvalues $\lambda_n^{\pm}$ satisfy $\lambda_n^{+}\lambda_n^{-} =1$ and are either 
complex conjugated with unit modulus ($\lambda_n^{\pm}= e^{\pm i\alpha_n}$), or real ($\lambda_n^{\pm}= e^{\pm \beta_n}$).
The first case ($\lambda_n^{\pm}= e^{\pm i\alpha_n}$) leads to stable solutions of the Hill equation of the general form
\beq
u_n(t)=C_n e^{i\gamma_n t}P_n(t)+{\tilde C}_n e^{-i\gamma_n t}{\tilde P}_n(t).
\label{un}
\eeq
Here $P_n(t)$ and ${\tilde P}_n(t)$ are complex-conjugated periodic functions with the same period $T$ as $a(t)$, $C_n$ and ${\tilde C}_n$ are arbitrary complex  coefficients, and   $\gamma_n \equiv\alpha_n/T $ is  real (and, say, positive: $\gamma_n>0$).  
In the unstable case ($\lambda_n^{\pm}= e^{\pm \beta_n}$), the general solution of Eq.~(\ref{Hill}) can still be written in the
form \eqref{un}, but with the understanding that $\gamma_n \equiv -i \beta_n/T $ be purely imaginary. In that case, the two
periodic functions $P_n(t)$ and ${\tilde P}_n(t)$ are both real.

We recall that the function $u_n(t)$, entering the representation \eqref{Snvsun} (and satisfying the Hill equation Eq.~(\ref{Hill}))
is just an auxiliary function that we introduced to solve the Riccati equation \eqref{riccatiallowed}. We must find some solution 
$u_n(t)$ such that the corresponding function $S_n=-i \frac{\dot u_n}{u_n}$ satisfies two different requirements:
(i) $S_n^{\pm}(a)$ must satisfy the matching conditions \eqref{matching1}, \eqref{matching2}, together with our strong
outer matching condition \eqref{matching3}; and (ii) they must satisfy (for all values of $a$) the stability inequality \eqref{regularity}.

If we did not have the extra outer matching condition \eqref{matching3}, we might have considered the possibility of constructing
a sequence of functions $S_{n N}^{\pm}(a)$ entering various branches of a wave function of the form
$ \Psi(a, \phi_n) = \sum_N  \Psi_N(a, \phi_n)$, with $\Psi_N(a, \phi_n)$ containing an $N$-dependent factor 
$\exp\left[ -\frac{1}{2}S_{ n N}^{\pm}(a) \phi_n^2 \right]$. Such a wave function might describe successive
cycles of oscillations of a quantum universe that would only differ in the quantum dispersion of the $\phi_n's$.
However, our outer matching condition \eqref{matching3}, in conjunction with the matching conditions \eqref{matching1}, \eqref{matching2},
require  $S_n(t)$ to be a {\it periodic function} of the Minkowskian time $t$, {\it i.e.} $S_{n (N+1)}^{+}(a)$ to be equal
to $S_{n N}^{+}(a)$, and $S_{n (N+1)}^{-}(a)$ to be equal
to $S_{n N}^{-}(a)$.
Indeed, suppose we follow the wave function (\ref{psiallowed}) for one oscillation period, as $a$ changes from its
maximum value $a_2$ to $a_1$ and back. We must start from a given value  $S_n^-(a_2) = \mu_n$. In the first half-period $S_n^-(a_2)=\mu_n$ evolves into some corresponding uniquely defined $S_n^-(a_1)$, which must equal  $S_n^+(a_1)$ in view of the junction condition (\ref{matching1}). Then, in the second half-period $S_n^+(a_1)=S_n^-(a_1)$ evolves into some corresponding $S_n^+(a_2)$.
The latter value must, in view of the other junction condition (\ref{matching2}), be equal to $S_n^-(a_2) $,
which must  (in view of the fixed outer matching condition \eqref{matching3})
be again equal to the same value $S_n^-(a_2) = \mu_n$, with which we started the cycle. Q.E.D.

 The mode functions $u_n(t)$ are not generally periodic, and, because of the multiplicative change of the eigenmodes after one period,
 their logarithmic derivative $ S_n=-i \frac{\dot u_n}{u_n}$ is also not generally periodic. It is easily seen that the only way to construct
a periodic $S_n(t)$ is to use just one of the two eigenmodes in Eq.~(\ref{un}). Using only the first mode, {\it i.e.} setting ${\tilde C}_n \to 0$ in Eq. \eqref{un}, yields
\beq \label{Snallowed}
S_n(t) = \gamma_n -i\frac{{\dot P}_n(t)}{P_n(t)}.
\eeq
This yields a periodic $S_n(t)$ both in the unstable case ($\gamma_n \equiv -i \beta_n/T $, with real $P_n(t)$), and in the
stable case ($\gamma_n \equiv\alpha_n/T >0$ with a complex periodic function $P_n(t)$). 
However, in the unstable case, the so-constructed periodic $S_n(t)$ is purely imaginary, and therefore does not satisfy the
regularity condition (\ref{regularity}), expressing that the $\phi_n$ wave function is localized near $\phi_n=0$. 
 We thus reach the physically reasonable conclusion that stationary states of an oscillating universe do not exist in the presence of unstable scalar field modes, {\it i.e.} in presence of particle creation effects. As said in the Introduction, we assume here that all the $\phi_n$
 modes are stable, therefore we continue our discussion by focussing on the stable case.

For a stable mode, $u_n(t)$, with $\gamma_n \equiv\alpha_n/T >0$, the construction \eqref{Snallowed} leads (for general values
of $a$) to a complex-valued
$S_n(t)$ (because $P_n(t)$ is a complex, periodic function) whose real part oscillates periodically around  $\gamma_n >0$:
$\langle {\rm Re}[ S_n(t)] \rangle = \gamma_n$. However, we must ensure that the regularity condition $Re[ S_n(t)] >0$, Eq. \eqref{regularity}, is satisfied not only on average, but at any time $t$ during an entire oscillation period.
It is, however, easy to show that this is automatically the case.

Indeed, let us decompose the complex number $u_n(t)$ in modulus and phase,
\beq
u_n(t)=\rho(t)e^{i\theta(t)}
\label{rhotheta}
\eeq
with $\rho$ and $\theta$ real.  Then
\beq
S_n={\dot\theta}-i\frac{\dot\rho}{\rho},
\label{Snrhotheta}
\eeq
so that
\beq
{\rm Re}[ S_n]={\dot\theta}.
\eeq
Substituting (\ref{rhotheta}) in Eq.~(\ref{Hill}) we obtain
\beq
{\ddot \rho}-\rho{\dot\theta}^2 +\omega_n^2 \rho +i(\rho{\ddot\theta}+2{\dot\rho}{\dot\theta})=0.
\eeq
Taking the imaginary part of this equation yields
\beq \label{conservation}
\frac{d}{dt}(\rho^2{\dot\theta})=0~~ {\rm or} ~~\rho^2{\dot\theta}={\rm const}.
\eeq
From Eq.~(\ref{conservation}), we see (barring singular cases were $\rho$ might vanish or become infinite) 
that  ${\rm Re}[S_n]={\dot\theta}$ will always keep the same sign
during an entire oscillation period. As its average $\gamma_n$ has been chosen to be
positive, we conclude that the stability condition \eqref{regularity} will remain satisfied all over the classically allowed range. 
Note in passing that Eq.~(\ref{conservation}) is actually just Kepler's area law, applied to the Newtonian dynamical
equation \eqref{Hill} for the motion of a particle moving in the complex plane under the influence of a time-dependent central force.

Summarizing so far: the matching conditions \eqref{matching1}, \eqref{matching2},\eqref{matching3}, impose
the periodicity of $S_n$ as a function of $t$, and  there is a {\it unique solution} of Eq. \eqref{riccatiallowed} in the
classically allowed range $a_1<a<a_2$ that is periodic,
and satisfies the  condition \eqref{regularity}). It is given by Eq. \eqref{Snallowed} (with $\gamma_n>0$).

However, that's where the good news end. Indeed, we must satisfy not only periodicity of $S_n(t)$, but also
the precise boundary condition $S_n(a_2)= \mu_n$, where the sequence of characteristic numbers $\mu_n>0$
is uniquely fixed by the structure of the function $\omega_n^2(a)$ (for $a>a_2$).

Let us first show that the values of $S_n(a_1)$ and $S_n(a_2)$, as obtained from Eq. \eqref{Snallowed}, are 
some, uniquely defined, {\it real} numbers, say
\beq
S_n(a_{(1,2)}) = \gamma_n + \delta_n^{(1,2)} \,.
\eeq
Indeed, the fact that the $t$-evolution of $a$ is defined by Eq. \eqref{xt}, {\it i.e.}
\beq \label{adot2}
\left(\frac{d a}{d t}\right)^2 = -U(a)\,,
\eeq
implies that we have time-reversal symmetry under $t\to t'= -t + {\rm const.}$. Actually, a moment of thought shows that the function
$a(t)$ (and thereby any function constructed from $a(t)$) is {\it even under time-reversal} around {\it both}
turning points. If we choose the origin of $t$ such that $a(0)=a_1$, we have $a(-t)=a(t)$. But, similarly,
if we shift $t$ into $t^{\rm new} = t - \frac12 T$ so that the origin of $t^{\rm new}$ corresponds to
$a(0^{\rm new})=a_2$, we also have $a(-t^{\rm new})=a(t^{\rm new})$.
The potential in Hill's equation \eqref{Hill} inherits these two symmetries. Therefore, the ($a_1$ and $a_2$) time reversals
of the first Floquet solution, $e^{i\gamma_n t}P_n(t)$, must be equal to the second Floquet solution, 
$e^{-i\gamma_n t}{\tilde P}_n(t)$, so that (using a star to denote complex conjugation)
\beq
P_n(-t)= {\tilde P}_n(t)= P_n^*(t)\,.
\eeq
Applying this result to both turning points, $t=0$ and $t^{\rm new}=0$, we have
\beq
P_n(0)= P_n^*(0)  ~~{\rm and}~~ {\dot P}_n(0)= - {\dot P}_n^*(0)  \,.
\eeq
In words: $P_n(a_j)$ is real, and $\dot P_n(a_j)$ is pure imaginary at {\it both} turning points, $j=1,2$,
so that
\beq
\delta_n^{(j)} \equiv - i \frac{\dot P_n(a_j)}{P_n(a_j)}
\eeq
are two real numbers, uniquely defined by the function $\omega_n^2(t)$.

The bad news is, however, that, as we have explicitly checked on some numerical examples,
no miracle occurs, and the uniquely defined outer boundary condition ${\bar S}_n^{\rm outer}(a_2)=\mu_n$ at $a_2$ {\it
generally  differs}
from the (uniquely defined) necessary, periodic inner boundary condition:
\beq \label{neqa2}
S_n^{\rm inner \; periodic}(a_{2}) = \gamma_n + \delta_n^{(2)}  \neq \mu_n = {\bar S}_n^{\rm outer}(a_2)\,.
\eeq
A simple case where we checked the non equality \eqref{neqa2} is an oscillatory universe filled with (positive) radiation,
together with a negative (fermioniclike \cite{Damour}) component $-C_F/a^6$, {\it i.e.} (with $C_r>0$ and $C_F >0$)
\beq
\rho(a) = + \frac{C_r}{a^4} - \frac{C_F}{a^6}\,,
\eeq
so that (with $k=+1$)
\beq
U(a)= k \,a^2 - C_r + \frac{C_F}{a^2}\,.
\eeq
It is easily seen that Eq. \eqref{adot2} yields a harmonic oscillator equation for the squared scale factor,
so that $a^2(t)= \alpha + \beta \cos (2 \pi \frac{t}{T})$, say with $0< \beta< \alpha$.
As a consequence, the potential $\omega_n^2(t)$ in Hill's equation is varying sinusoidally for all modes numbers $n$:
\beq
 \omega_n^2(t)= n^2 + m^2 \alpha + m^2 \beta \cos \left(2 \pi \frac{t}{T}\right)\,.
 \eeq
 In other words, the Hill equations \eqref{Hill} reduce to Mathieu's equation (and to their Euclidean avatars,
 with $\omega_n^2(\tau)= n^2 + m^2 \alpha \pm m^2 \beta \cosh \left(2 \pi \frac{\tau}{T}\right)$ in the forbidden regions
 $a >a_2$, or $a< a_1$, respectively).  Using then both the analytical knowledge on Mathieu's solutions, and a sample of numerical
 simulations, we have verified the generic absence of equality \eqref{neqa2}.

\subsection{Quasi-uniqueness of the solution for $S_n(a)$ in the inner classically forbidden region $a<a_1$, and its
failure to match the $S_n(a)$'s in the other intervals.}

At this stage, we have already proven the impossibility to find stable quantum states of our system. It is, however, interesting
to complete our analysis by considering the inner classically forbidden region  $a<a_1$. This will show the existence of
further impossibilities. 
 
 It is again convenient to define a Euclidean time variable $\tau$ by
\beq 
\frac{da}{d\tau}=S'(a)=- \sqrt{U(a)} ~~~~ (a <a_1).
\label{atau}
\eeq
This Euclidean time grows when $a (<a_1)$ varies between $a_1$ and 0, and can be set to be zero when $a=a_1$.
The situation is then symmetric to our discussion of the $a> a_2$ region, with $\tau$ growing from 0 to some
(finite or infinite) maximum value $\tau_m$. The equations to solve are exactly the same as those in the outer classically forbidden
domain, namely the Riccati equation \eqref{riccatiforbidden}, and its linear second-order transform \eqref{untaueq}, obtained by
representing $S_n(\tau)$ in the form \eqref{Snuntau}.

The only difference with the discussion in the previous subsection concerns the total number
of e-folds
\beq
N_1=\int_{0}^{\tau_m} \omega_n(\tau)d\tau\,,
\eeq
 that will enter the approximate WKB solution (where $D_{n \pm} ={\rm const}$)
\beq
u_n(\tau)\approx   D_{n +}\exp\left[\int^\tau_0 \omega_n(\tau')d\tau'\right] + D_{n - }\exp\left[-\int^\tau_0 \omega_n(\tau')d\tau'\right]\,,
\label{untau1}
\eeq
when considering its limit as $\tau \to \tau_m$, {\it i.e.} as $a \to 0$. [As in the previous subsection, one could work with
the exact solutions corresponding to $\exp \pm \int^\tau_0 \omega_n(\tau')d\tau'$ and reach the same conclusions,
as long as $N_1$ is infinite or at least sufficiently large.]
The explicit expression of $N_1$ reads
\beq
N_1=\int_0^{a_1} da \frac{ \sqrt{n^2+ m^2 a^2}}{\sqrt{U(a)}} = \int_0^{a_1} da \frac{ \sqrt{n^2+ m^2 a^2}}{\sqrt{a^2[1- a^2 \rho(a)]}} .
\eeq

Let us first assume that $a^2 \rho(a) \to 0$ (or at least $ < 1$) as $a\to 0$ (this is notably the case for the SHU model \cite{Graham:2011nb}). Then  $U(a)\approx a^2$ as $a\to 0$ so that both $N_1$ and $\tau_m = \int_0^{a_1} da /{\sqrt{U(a)}}$ are
  logarithmically infinite.
In this case, we can rigorously reach the same conclusion as in the discussion of the $a>a_2$ region. 
Any nonzero value for the constant $ D_{n +}$ leads
to the asymptotic behavior
\beq
u_n(\tau)\approx  D_{n +} \exp\left[\int^\tau_0 \omega_n(\tau')d\tau'\right] \sim D_{n +} \exp n \,\tau ~~~{\rm as} ~~\tau \to \infty\,,
\eeq
which implies
\beq
S_n(a\to 0) \approx -\omega_n(a \to 0) \sim  - n  <0 ~~~~ ({\rm if} ~~ D_{n +}\neq 0).
\eeq
Therefore, there is a {\it unique} solution of the Riccati equation \eqref{riccatiforbidden} for $S_n(a)$ in the domain $a<a_1$
that satisfies the stability requirement \eqref{regularity}. In the WKB approximation it is given by setting $D_{n +} = 0$ in
Eq. \eqref{untau1}, and is approximately given by
\beq
S_n(a) \approx +{\omega_n(a)} > 0 ~~~~ ({\rm if} ~~ D_{n +} = 0).
\eeq

Like in the $a>a_2$ case, the reasoning of Ref. \cite{Messiah} shows the existence of an exact counterpart of this
solution (with $S_n(a)$ not exactly equal to, but close to $+{\omega_n(a)}$), constructed as   
$ S_n(a) =   - \dot u_n^-/u_n^- $ where $u_n^- (\tau)$ is the unique (positive, monotonously
decreasing, and asymptotically vanishing) solution of Eq. \eqref{untaueq} (normalized to equal 1 at $\tau=0$).

Suppose now that, as $a\to 0$, $U(a)$ either decreases slower than $a^2$ or even increases. [This would be the case if, {\it e.g.}, 
$a^2 \rho(a) \to - \infty$ as $a \to 0$, as it happens for models including a term of the type $\rho_F(a) = - C_F/a^6$.] 
Then both $\tau_m$ and $N_1$ are finite.  In that case, the inequality \eqref{regularity} is not powerful
enough to select a mathematically unique solution. However, as 
\beq
\omega_n(a) = \sqrt{n^2+ m^2 a^2} \approx n ~~~{\rm as} ~~ a \to 0,
\eeq
the number  $N_1$ of e-folds will grow linearly with the mode number $n$, and will therefore be very large
for an infinite number of modes. As a consequence, there will be a quasi-uniqueness of the solutions 
satisfying \eqref{regularity}, in the sense that for an infinite number of values of $n$ one must fine-tune 
the only integration constant $ D_{n + }/D_{n -}$  entering $S_n(a)$ to be increasingly closer to zero, as $n$ increases.

And again, we have verified that the so-determined boundary condition for $S_n(a_1)$, namely
\beq
{\bar S}_n^{\rm inner}(a_1) = \left[ - \frac{\dot u_n^-}{u_n^- }\right]^{a = a_1}
\eeq
will not match any would-be periodic allowed-region solution (which must have ${\bar S}_n^{\rm inner}(a_1) = \gamma_n + \delta_n^{(1)}$).

\section{Conclusions}

Let us recap our logic and our main results.
We investigated stationary quantum states of a compact universe filled with homogeneous matter and a massive scalar field conformally coupled to the curvature.  The background value of the scalar field was taken to be vanishing, 
and the quantum scalar field was accordingly treated as a perturbation.  The matter energy density $\rho(a)$ was assumed to be such that the unperturbed model has a classically allowed range for the scale factor, $0<a_1 <a<a_2$, with the values of $a$ beyond this range being classically forbidden.  Our main assumption is that the wave function vanishes at $a\to\infty$; 
we also tentatively required that the wave function also vanishes at $a=0$, but this is not necessary for reaching our
main results. The unperturbed wave function describes a universe oscillating periodically between $a_1$ and $a_2$.  We required that 
the usually considered source of quantum instability of an oscillating universe (particle creation due to parametric amplification
of the modes of $ \phi({\bf x},t)=\frac{1}{a(t)}\sum_n \phi_n(t) Q_n({\bf x})$) be absent.

If a stationary state exists, its full wave function should be given by a linear combination of Gaussian terms of the form 
\beq
\Psi (a,\{\phi_n\})=\exp\left[-\frac{S(a)}{\hbar}-\frac{1}{2\hbar}\sum_n S_n(a) \phi_n^2\right],
\label{Gaussiansbi}
\eeq
 with the functions $S_n(a)$ satisfying the stability condition ${\rm Re}[ S_n(a)] >0$ in the three intervals
 $0<a<a_1$, $a_1 <a<a_2$, and $a>a_2$, together with the matching conditions \eqref{matching1}, \eqref{matching2}.
 This condition ensures that the dependence on the amplitudes of the scalar field modes $\phi_n$ is Gaussian, rather than inverse Gaussian, so that the wave function does not grow with increasing $|\phi_n|$.  We found, however, that the stability condition cannot be satisfied in the full range of
variation of $a$, and is generally violated for an infinite number of modes.  

With a suitable choice of boundary conditions,
 the  stability condition ${\rm Re}[ S_n(a)] >0$  can be satisfied in one of the three intervals, $0<a<a_1$,
 $a_1 <a<a_2$, or $a>a_2$, but then it is violated in the two other intervals.
 Our conclusion is particularly sharp if we start imposing the stability condition in the outer classically forbidden range $a> a_2$.
 This determines, in a unique manner, the $S_n(a)$ function in the domain $a>a_2$. The matching conditions at the turning points $a_2$
 and $a_1$ then determine $S_n(a)$ in the other intervals. However, this continuation to $a<a_2$ does not lead to a
 consistent, and stable solution. [If one tries  to continue $S_n(a)$ in the classically allowed range starting from 
 the unique $S_n(a_2)=\mu_n$, one ends up, after one cycle of oscillation, with an inconsistent new (complex)
 value of $S_n(a_2)$. Moreover, the value of $S_n(a_1)$ obtained after the first half cycle would also be complex
 and inconsistent with stability in the $a<a_1$ domain.]
 
 The conclusion is thus that stationary quantum states for an oscillating universe do not exist.  We have verified that a similar analysis goes
  through, with the same conclusions, for a minimally coupled scalar field.

Our analysis was very general: we did not assume any specific form of the matter density function $\rho(a)$, requiring only that it should yield a classically allowed range of $a$ flanked by two classically forbidden ranges.  One may be concerned that the same kind of analysis could be used to demonstrate non-existence of stationary states in ordinary quantum mechanics.  There is, however, an important difference between the two cases.  The WDW equation is hyperbolic, while the stationary Schr\"odinger equation is elliptic, with all derivative terms having the same sign.  By applying our type of analysis to the case where the sign of the kinetic term for the scale factor is changed (with a corresponding
change in the confining potential $U(a)$), {\it i.e.} for a Schr\"odinger equation of the type
\beq
\left[-\hbar^2\frac{\partial^2}{\partial a^2} +U(a) +\sum_n\left(-\hbar^2\frac{\partial^2}{\partial \phi_n^2} +\omega_n^2(a) \phi_n^2 \right)\right] \Psi =0,
\label{WDWEuclidean}
\eeq
we have shown that our method meets no obstacle for constructing stable solutions, with ${\rm Re}[ S_n(a)] >0$.
The crucial difference is that we find that the sign of the term $S' {S_n}'$ in Eq. \eqref{Sneq} is reversed, which implies that
one must correspondingly reverse the signs of the logarithmic derivatives on the right-hand-sides of the representations 
\eqref{Snvsun} and \eqref{Snuntau} (the other equations remaining the same). This apparently minor sign change has
a drastic effect on the selection of the stable solutions in the two forbidden domains. Indeed, when having 
$S_n=+\frac{{\dot u}_n}{u_n}$, the naturally dominating growing Euclidean solutions 
$u_n^+(\tau)\approx  C_+ \exp\left[\int^\tau_0 \omega_n(\tau')d\tau'\right]$ yield the stable value 
$S_n(a) \approx +\omega_n(a) $, so that no fine tuning of the integration constants is needed to select a stable solution.
One can then start (using Floquet theory) by constructing a stable, periodic function $S_n(a)$ in the domain $a_1<a<a_2$.
This defines positive, real boundary values $S_n(a_{(1,2)})= \gamma_n + \delta_n^{(1,2)}$, which can then
be continued into stable solutions in the two classically forbidden intervals $ a<a_1$ and $a> a_2$.

Our result appears to point to some quantum instability of an oscillating universe. 
Mathematically, it is rooted in the fact that our WDW equation \eqref{WDW} is a
Klein-Gordon equation,  $ ({\widehat P}^2 + M^2) \Psi=(- \hbar^2\Box + M^2) \Psi =0$ (with signature $- ++++\cdots$), 
where $a$ is the time variable, and in which the squared mass is the following function on superspace:
\beq
M^2(a, \phi_n) =  -U(a) +\sum_n\omega_n^2(a) \phi_n^2 \,.
\eeq
This squared mass is positive in a connected domain of the variables $(a,\phi_n)$ which is centered along the line $(a_1<a<a_2, \phi_n=0)$
and extends away from it in all the $\phi_n$ directions (and also in part of the
regions where $a<a_1$ or $a>a_2$, if $|\phi_n|$ is large enough),
but it is {\it negative} in two other (disconnected)  domains: an infinite region centered along the infinite half-line $( a>a_2, \phi_n=0)$
and extending around it in the $\phi_n$ directions, and a  third region centered along the interval $( 0<a<a_1, \phi_n=0)$
and extending around it in the $\phi_n$ directions. In the latter two domains (centered around the classically forbidden lines,
$(0<a<a_1, \phi_n=0)$ and $( a>a_2, \phi_n=0)$), the quantum particle-universe described by the WDW equation
is a {\it tachyonic} particle, which gives rise to strong, exponential instabilities in its quantum propagation in superspace.
These tachyonic exponential instabilities are the mathematical roots of the instabilities we found in our analysis.

The physical nature of these instabilities remains unclear.  Rubakov {\it et al} \cite{Rubakov1,Rubakov2} found a similar instability in the under-barrier wave function of a tunneling universe.  They argued that particles must be copiously ("catastrophically") produced in the course of tunneling.  An alternative explanation suggested in Refs.~\cite{Hong1,Hong} is "critical branching", meaning that the  universe is a superposition of 
different quasi-classical branches, some of which 
contain a larger number of particles, leading to a smaller barrier that needs to be penetrated.
The wave function after tunneling may then be dominated by states with high occupation numbers, even if such states are strongly suppressed in the initial wave function.  Furthermore, it was shown in \cite{Hong} that the initial state can always be fine-tuned to remove the multi-particle branches, so that the wave function exhibits no instability.

This situation is different from the case of an oscillating universe, where the instability is present for any choice of the quantum state.  It is still possible that our result is a reflection of critical branching, which now cannot be removed by a choice of the quantum state because it is present in {\it two} under-barrier regions rather than one.  On the other hand, even though the effect of critical branching is certainly present, it is not clear that it is accounted for in our formalism, which seems to disregard the effect of particle occupation numbers on the expansion of the universe.  So it is conceivable that our result is a reflection of an instability due to some nonperturbative particle production mechanism.  This issue remains an intriguing problem for future research.

\section*{Acknowledgments} 
T.D. thanks Gerald Dunne, Jens Hoppe, Sergiu Klainerman, Marcos Mari\~no, Vasily Pestun, Valery Rubakov and Andr\'e Voros
 for informative discussions. A.V. is grateful to the Institut des Hautes Etudes Scientifiques for its hospitality 
 during the conception of this work. A.V. acknowledges support by the National Science Foundation under grant
 PHY-1820872.

\end{document}